\documentclass{article}
\usepackage[margin=25mm]{geometry}
\usepackage[hyphens]{url}
\usepackage{listings, listing}

\usepackage{amssymb}
\usepackage{amsmath}
\usepackage{amsthm}
\usepackage{color}
\usepackage{braket}
\usepackage{graphicx}
\usepackage{bbm}

\newtheorem{theorem}{Theorem}
\newtheorem{lemma}[theorem]{Lemma}

\newcommand{\select}{\mathsf{select}}

\newcommand{\ZZ}{\mathbb{Z}}

\newcommand{\ignore}[1]{}

\def\qed{\hfill $\Box$}

\newcommand{\ceil}[1]{\lceil #1 \rceil}
\newcommand{\floor}[1]{\lfloor #1 \rfloor}

\newcommand{\zdiv}{\ensuremath{\mathbin{/\mkern-4mu/}}}
\newcommand{\idiv}[2]{#1 {\zdiv} #2}
\newcommand{\imod}[2]{#1 {\rm \% } #2}

\newcommand{\tablecap}[2]{{
\begin{table}[htbp]
\begin{center}
\caption{#1}
#2
\end{center}
\end{table}
}}

\lstdefinelanguage{asm}{
  morecomment=[l]{;},
  morecomment=[l]{\#},
  morestring=[b]",
  sensitive=true
}

\newcommand{\lstinputlistingasm}[2][]{%
  {\catcode`\%=12\relax\lstinputlisting[mathescape=false,#1]{#2}}%
}

\begin{document}

\title{
  Optimization of 32-bit Unsigned Division by Constants on 64-bit Targets
}
\author{
    Mitsunari Shigeo
    \thanks{
        Cybozu Labs, Inc.
    }
    \and
    Hoshino Takashi
    \footnotemark[1]
}
\date{}
\maketitle

\begin{abstract}
Granlund and Montgomery proposed an optimization method for unsigned integer division by constants \cite{GM}.
Their method (called the GM method in this paper) was further improved in part by works such as \cite{CDW} and \cite{Hacker}, and is now adopted by major compilers including GCC, Clang, Microsoft Compiler, and Apple Clang.
However, for example, for x/7, the generated code is designed for 32-bit CPUs and therefore does not fully exploit 64-bit capabilities.

This paper proposes an optimization method for 32-bit unsigned division by constants targeting 64-bit CPUs.
We implemented patches for LLVM/GCC and achieved speedups of 1.67x on Intel Xeon w9-3495X (Sapphire Rapids) and 1.98x on Apple M4 (Apple M-series SoC) in the microbenchmark described later.
The LLVM patch has already been merged into llvm:main \cite{LLVMopt}, demonstrating the practical applicability of the proposed method. 

\end{abstract}

\section{Notation}
We use the following notation.
Let $\ZZ$ be the set of integers. For $a, b \in \ZZ$, define $[a, b] := \Set{c \in \ZZ \mid a \leq c \leq b}$.
For a real number $x$, define $\floor{x} := \max \Set{y \in \ZZ \mid y \leq x}$ and $\ceil{x} := \min \Set{y \in \ZZ \mid y \geq x}$.
For integers $a \geq 0$ and $b \geq 1$, let $q$ and $r$ be the quotient and remainder of dividing $a$ by $b$: $q := \idiv{a}{b} = \floor{a/b}$, $r := \imod{a}{b}$.
That is, $a = bq + r$ and $0 \leq r < b$.
For nonnegative integers $x, a$, define logical right shift by $x \gg a := \idiv{x}{2^a}$.
Let nonnegative integer $M$ be the maximum possible dividend and let divisor $d \in [1, M]$.
Define $\select({\tt cond}, x, y)$ as the function that returns $x$ when condition {\tt cond} is true and returns $y$ otherwise.
We denote 32/64/128-bit unsigned integer types by {\tt u32/u64/u128}.

\section{The GM Method}
We first review existing approaches.
Fix integers $M \geq 1$ and $d \geq 1$, and consider $\idiv{x}{d}$ for $x \in [0, M]$.
The following theorem is used by optimizing compilers such as GCC and Clang \cite{GM}\cite{Hacker}\cite{Optimal}.
\begin{theorem}[GM-method]\label{TH_GM}
Fix $d \in [1, M]$.
Let $M_d := \max \Set{x \in [0, M] \mid \imod{x}{d}=d-1}$.
For positive integers $c, A$, if $1/d \leq c/A < (1+1/M_d)/d$, then $\idiv{x}{d}=\idiv{(xc)}{A}$ holds for all $x \in [0, M]$.
\end{theorem}

For a given $A$, let $c := \ceil{A/d}$ and $e := cd - A$.
Then $0 \leq e < d$ and $1/d \leq c/A$.
The condition $c/A < (1+1/M_d)/d = (M_d+1)/(dM_d)$ is equivalent to $cdM_d < (M_d+1)A$ after multiplying both sides by $dM_d$, which is equivalent to $eM_d < A$.

In the rest of this section, we focus on 32-bit unsigned integers and set $M = 2^{32}-1$.
When optimizing $\idiv{x}{d}$, compilers can be classified into four major patterns depending on $d$.

\begin{lstlisting}[
    language=C++,
    caption={u32 constant division},
    label={CODE_ref},
]
u32 udivd(u32 x) {
  // d is a global u32 constant.
  return x / d;
}
\end{lstlisting}

\subsection{Case: $d$ is a power of two}
If $d=2^a$ (integer $a$), then $\idiv{x}{d} = x \gg a$ and division is replaced by a logical right shift.

\subsection{Case: $d>\idiv{M}{2}$}
Since $\idiv{x}{d} \in \Set{0,1}$, it can be implemented as $\idiv{x}{d} = \select(x<d, 0, 1)$.

\subsection{Other cases}
\begin{lemma}
If $M$ is an $m$-bit integer and $d$ is not a power of two, then $c$ can be chosen to be at most $m+1$ bits.
\end{lemma}
Since $M$ is $m$ bits, $2^{m-1} \leq M < 2^m$.
Let $d$ be a $b$-bit integer that is not a power of two, then $2^{b-1} < d < 2^b$.
Because $e<d$, we have $eM_d < dM < 2^{m+b}$.
Hence if we choose $A=2^{m+b}$, the theorem condition is satisfied.
Then $c=\ceil{A/d} \leq 2^{m+b-b+1}=2^{m+1}$.
If $c=2^{m+1}$, we can replace $(c, A)$ with $(c/2, A/2)$ while preserving the theorem condition, so we can make $c=2^m$.
Therefore, $c<2^{m+1}$ can always be achieved.
\qed

In particular, for $m=32$, $c$ is at most 33 bits.

When $c$ fits in 32 bits, a standard implementation is
$\idiv{x}{d}=(xc)\gg a$,
i.e., multiply and then logically right-shift the 64-bit result.
The 33-bit $c$ case is the main target of this paper.
An exhaustive checking over all 0x7fffffff nontrivial values of d ($d<0x80000000$) shows that about 77\% fit in 32 bits and about 23\% require 33 bits.

\subsection{Case: $c$ is 33 bits}
At the time of our verification (March 2026), assembly outputs for x86-64 and Apple M-series generated by GCC (14.2.0), Clang (18.1.3), Microsoft Compiler in Visual Studio 2026 (19.50.3571), and Apple clang (17.0.0) all use an algorithm equivalent to Listing \ref{CODE_gm}.
For x86-64, the comparison used \texttt{-O2 -mbmi2}.

\begin{lstlisting}[
    language=C++,
    caption={GM method when c is 33 bits},
    label={CODE_gm},
]
u32 udivd(u32 x) {
  // c and a are constants determined by d.
  u32 cL = c & 0xffffffff;
  u32 y = (u64(x) * cL) >> 32;
  u32 t = ((x - y) >> 1) + y;
  return t >> (a - 33);
}
\end{lstlisting}

We explain Listing \ref{CODE_gm}.
When $c$ is 33 bits, $a \geq 33$, and
$$(xc)\gg a=(((xc)\gg 32)\gg 1)\gg (a-33)$$
can be rewritten as three shifts.
Let $c_L$ be the low 32 bits of $c$, then $c=2^{32}+c_L$ and
$xc=x(2^{32}+c_L)=x2^{32}+xc_L$.
Let $y := (xc_L)\gg 32$, then $(xc)\gg 32 = y+x$.

Using $y=\idiv{(xc_L)}{2^{32}}\leq\idiv{(x2^{32})}{2^{32}}=x$, we get
$2^{32}>\idiv{(y+x)}{2}=\idiv{(x-y+2y)}{2}=\idiv{(x-y)}{2}+y$.
This transformation avoids intermediate values exceeding 32 bits.

For example, Listing \ref{CODE_div7_m4_org} is the assembly output for $d=7$ when Listing \ref{CODE_ref} is compiled with {\tt -Ofast} (parameters: $a=35$, $c={\tt 0x124924925}$).

\begin{lstlisting}[
    language=C++,
    caption={Apple M4 Clang assembly output},
    label={CODE_div7_m4_org},
]
; input: x = w0
; output: w0
; d = 7
; a = 35
udiv:
  ; cL = w8 = 0x24924925
  mov w8, #18725  ; =0x4925
  movk    w8, #9362, lsl #16
  ; x8 = cL * x
  umull   x8, w0, w8
  ; y = x8 = (cL * x) >> 32
  lsr x8, x8, #32
  ; w9 = x - y
  sub w9, w0, w8
  ; t = ((x - y) >> 1) + y
  add w8, w8, w9, lsr #1
  ; t >> (a - 33)
  lsr w0, w8, #2
  ret
\end{lstlisting}

\section{Proposed Method}
Listing \ref{CODE_gm} is designed so that intermediate values (except for multiplication itself) fit within 32 bits.
On 64-bit CPUs, this constraint is no longer necessary.
A straightforward approach would be to directly compute the {\tt u64}×{\tt u64}={\tt u128} multiplication and then right-shift the result.
However, this is inefficient on x86-64 architectures because the 128-bit logical right-shift instruction {\tt shrd} requires the same latency and throughput as {\tt mul} on Skylake-X \cite{anger}.
A similar trend was observed on Sapphire Rapids used in our benchmark.

\tablecap{Skylake-X architecture: latency and throughput\cite{anger}}{
\label{TBL_xeon}
\begin{tabular}{|l|c|c|}
\hline
Instruction & Latency & Throughput \\\hline
add/sub & 1 & 0.25 \\\hline
shr & 1 & 0.5 \\\hline
imul & 3 & 1 \\\hline
shrd & 3 & 1 \\\hline
\end{tabular}
}

Therefore, we propose the following optimization.
$$\idiv{(xc)}{2^a}=\idiv{(x(2^{64-a}c))}{2^{64}}$$
where $x$ is zero-extended to 64 bits.
Since $c$ is 33 bits and $d \le \idiv{M}{2}$, we have $33 \le a \le 63$, so $2^{64-a}c$ fits in 64 bits.
A 128-bit value is represented by two 64-bit registers $(H, L)$, so a 64-bit logical right shift is just extracting $H$.
Except for loading the immediate constant, only one multiplication instruction is needed.
On AArch64 architectures such as Apple M4, the {\tt u64}×{\tt u64}={\tt u128} multiplication is split into {\tt umulh}, which returns the upper 64 bits, and {\tt mul}, which returns the lower 64 bits.
This transformation allows the computation to be implemented with a single {\tt umulh} instruction.
The x86-64 form is Listing \ref{CODE_div_xeon_opt} and the M4 form is Listing \ref{CODE_div_m4_opt}.
In loops with repeated divisions by the same constant, the immediate is reused by optimization, so it is effectively one multiply instruction.

\begin{lstlisting}[
    language=C++,
    caption={Proposed method for x86-64},
    label={CODE_div_xeon_opt},
]
// input: x
// output: rax = x // d
// destroy: t
mov(t, c << (64 - a));
mulx(rax, t, x); // [rax:t] = x * t
\end{lstlisting}

\begin{lstlisting}[
    language=C++,
    caption={Proposed method for M4},
    label={CODE_div_m4_opt},
]
// input: x
// output: x0 = x // d
// destroy: t
mov_imm(t, c << (64 - a));
umulh(x0, t, x); // x0 = (t * x) >> 64
\end{lstlisting}

\section{LLVM/GCC Improvements and Benchmark}
We implemented the proposed optimization in LLVM (the compiler infrastructure of Clang) \cite{LLVMopt}.
To evaluate the effect, we prepared benchmark code for 32-bit unsigned division by constants.
In Listing \ref{CODE_bench}, divisors 7, 19, and 107 are cases where $c$ becomes 33 bits.

\begin{lstlisting}[
    language=C++,
    caption={bench.c},
    label={CODE_bench},
]
int test(uint32_t ret) {
  for (int i = 0; i < 1000000000; i++) {
    ret ^= (i ^ ret) / 7;
    ret ^= (i ^ ret) / 19;
    ret ^= (i ^ ret) / 107;
  }
  return ret;
}
\end{lstlisting}

First, we generated LLVM IR \texttt{bench.ll} with
\texttt{clang-18 -O2 -S -emit-llvm bench.c -o bench.ll}.
Then we used LLVM's \texttt{llc} to generate assembly for each CPU.

\noindent
For Xeon:
\texttt{llc -O2 -mattr=+bmi2 bench.ll -o <output>.s}.

\noindent
For M4:
\texttt{llc -O2 -mtriple=arm64-apple-macosx -mcpu=apple-m4 bench.ll -o <output>.s}.

We compare the code generated by the original LLVM 23.0.0 \texttt{llc} with the code generated by \texttt{llc} modified to support the proposed method.

\lstinputlistingasm[
    language=asm,
    caption={bench-x64-org.s},
    label={CODE_x64_org},
]{src/bench-x64-org.s}

\lstinputlistingasm[
    language=asm,
    caption={bench-x64-opt.s},
    label={CODE_x64_opt},
]{src/bench-x64-opt.s}

\lstinputlistingasm[
    language=asm,
    caption={bench-m4-org.s},
    label={CODE_m4_org},
]{src/bench-m4-org.s}

\lstinputlistingasm[
    language=asm,
    caption={bench-m4-opt.s},
    label={CODE_m4_opt},
]{src/bench-m4-opt.s}

Listings \ref{CODE_x64_org} and \ref{CODE_x64_opt} are for x86-64, and Listings \ref{CODE_m4_org} and \ref{CODE_m4_opt} are for Apple M4.
Comparing each pair shows that before the change, 33-bit constant division is implemented with a three-stage shift sequence,
whereas after the change it is implemented by a single multiply, resulting in a more compact loop.

Next, we assembled these files into executables and measured runtime with the time command.
Results are shown in Table \ref{TBL_bench}.

\tablecap{Runtime comparison on Xeon/Apple M4 (sec)}{
\label{TBL_bench}
\begin{tabular}{|l|r|r|r|}
\hline
CPU & Before & After & Speedup \\ \hline
Xeon & 6.33 & 3.80 & x1.67 \\ \hline
M4 & 6.70 & 3.38 & x1.98 \\
\hline
\end{tabular}
}

Measurement environments were Xeon w9-3495X (Sapphire Rapids)/Ubuntu 24.04.4 LTS and Apple M4 MacBook Pro/Tahoe 26.4.
Each measurement was repeated 10 times and the average was taken.
CPU settings were the default frequency/performance settings.
The average runtime improved from 6.33 sec to 3.80 sec (x1.67) on Xeon and from 6.70 sec to 3.38 sec (x1.98) on M4.
The sample standard deviations on Xeon were 0.013 sec (before) and 0.009 sec (after), indicating low variance.

This LLVM change has been merged into llvm:main.
We also prepared a corresponding GCC patch \cite{GCCpatch}, which is currently under review.

\section{Conclusion}
This paper proposed an optimization method for 32-bit unsigned division by constants under a 64-bit CPU assumption.
For the 33-bit-constant case in the conventional GM method, we showed that implementation is possible with a single multiplication instruction without a 128-bit shift.
With the LLVM implementation, we confirmed up to 1.67x (Sapphire Rapids) and 1.98x (Apple M4) speedups in real machine benchmarks.
We also prepared a corresponding GCC patch.

\bibliographystyle{abbrv}
\bibliography{main}

@ARTICLE{CDW,
  author={Cavagnino, D. and Werbrouck, A. E.},
  journal={The Computer Journal},
  title={Efficient Algorithms for Integer Division by Constants Using Multiplication},
  year={2008},
  volume={51},
  number={4},
  pages={470-480},
  doi={10.1093/comjnl/bxm082}
}

@book{Hacker,
  author       = {Warren, Jr., Henry S.},
  title        = {Hacker's Delight, Second Edition},
  publisher    = {Pearson Education},
  year         = {2013},
  url          = {http://www.hackersdelight.org/},
  isbn         = {0-321-84268-5},
  bibsource    = {dblp computer science bibliography, https://dblp.org}
}

@article{GM,
author = {Granlund, Torbjorn and Montgomery, Peter},
journal = {ACM SIGPLAN Notices},
volume = {29},
number = {6},
year = {1994},
pages = {61--72},
title = {Division by invariant integers using multiplication}
}

@article{Optimal,
  author       = {Daniel Lemire and
                  Colin Bartlett and
                  Owen Kaser},
  title        = {Integer Division by Constants: Optimal Bounds},
  journal      = {CoRR},
  volume       = {abs/2012.12369},
  year         = {2020},
  url          = {https://arxiv.org/abs/2012.12369},
  eprinttype    = {arXiv},
  eprint       = {2012.12369},
  bibsource    = {dblp computer science bibliography, https://dblp.org}
}

@misc{anger,
author={Anger Fog},
year={2025},
month={9},
title={Instruction tables},
note={\url{https://www.agner.org/optimize/instruction_tables.pdf}},
}

@misc{LLVMopt,
  author={Mitsunari Shigeo},
  year={2026},
  title={{[SelectionDAG] Optimize 32-bit udiv with 33-bit magic constants on 64-bit targets}},
  note={\url{https://github.com/llvm/llvm-project/pull/181288}},
}

@misc{GCCpatch,
  author={Mitsunari Shigeo},
  year={2026},
  title={optimize-udiv32-on-64bit},
  note={\par\noindent\url{https://github.com/herumi/gcc/commits/optimize-udiv32-on-64bit/}},
}

\end{document}